\documentstyle[preprint,epsfig,aps]{revtex}

\begin{document}

\title{Neutrinoless double beta decay within Self-consistent Renormalized
Quasiparticle Random Phase Approximation and inclusion of induced nucleon
currents}

\author{ A. Bobyk, W. A. Kami\' nski}
\address{Department of Theoretical Physics,
Maria Curie-Sk{\l}odowska University,\\
Radziszewskigo 10, PL-20-031 Lublin, Poland.}

\author{F. \v Simkovic}
\address{Department of Nuclear Physics, Comenius University,\\
Mlynska dol., pav. F1, SK-842 15 Bratislava, Slovakia}

\date{\today}
\maketitle
%%%%%%%%%%%%%%%%%%%%%%%%%%%%%%%%%%%%%%%%%%%%%%%%%%%%%%%%%%%%%%%%%%%%%%%
%                          ABSTRACT                                   %
%%%%%%%%%%%%%%%%%%%%%%%%%%%%%%%%%%%%%%%%%%%%%%%%%%%%%%%%%%%%%%%%%%%%%%%

\begin{abstract}
The first, to our knowledge, calculation of neutrinoless double
beta decay ($0\nu\beta\beta$-decay ) matrix elements within
the self-consistent renormalized Quasiparticle Random Phase Approximation
(SRQRPA) is presented. The contribution from the momentum-dependent
induced nucleon currents to $0\nu\beta\beta$-decay amplitude
is taken into account. A detailed nuclear structure study includes the
discussion of the sensitivity of the obtained SRQRPA results 
for $0\nu\beta\beta$-decay of $^{76}$Ge to
the parameters of nuclear Hamiltonian, two-nucleon short-range 
correlations and the truncation of the model space.
A comparison with the standard and renormalized QRPA results
is presented. We have found a considerable reduction of 
the SRQRPA nuclear matrix elements, resulting in less stringent limits 
for the effective neutrino mass. 
\end{abstract}

The neutrino mass and mixing of neutrinos are the main subject 
of the elementary particle physics nowadays. The 
experiments looking for oscillation 
of solar, atmospheric and  terrestrial (LSND-experiment)
neutrinos constitute evidence for a new physics beyond
the Standard Model \cite{zub}. The current constraints
imposed by the results of neutrino oscillation
experiments allow to construct the spectrum of mass
of neutrinos \cite{bil99}. The predictions differ from each other 
by different input, structure of neutrino mixing and
assumption, e.g., on the phases and fundamental character
of neutrinos. 

An important quantity to limit the space of 
possible neutrino mixing schemes is the effective Majorana
neutrino mass 
\begin{eqnarray}
\langle m_{\rm ee}\rangle = 
  \left| \sum_k~ (U_{{\rm e}k})^2 m_k \eta^{\rm CP}_k \right|
\label{eq:1}   
\end{eqnarray}
where $U_{{\rm e}k}$, $m_k$ and $\eta_k^{\rm CP}$ are  unitary mixing matrix
elements, mass eigenstates and the relative CP-phases of neutrinos,
respectively. 
The value of $\langle m_{\rm ee}\rangle$ depends on the specific model of 
neutrino mixing and its predictions fall in the range
of $10^{-3}$ eV to few eV \cite{bil99}. 
The upper bound on $\langle m_{\rm ee}\rangle$ can be deduced from the
current experimental lower bound on the half-life of
neutrinoless double beta decay ($0\nu\beta\beta$-decay) 
as follows \cite{doi85,hax,fae98}:
\begin{equation}
{\langle m_{\rm ee}\rangle} \le \frac{m_{\rm e}}{|M^{}_{\langle m_{\rm ee}\rangle}|
\sqrt{ T_{1/2}^{0\nu-{\rm exp}} G_{01}}}. 
\label{eq:2}   
\end{equation}
Here, $m_{\rm e}$, $G_{01}$ and $M^{}_{\langle m_{\rm ee}\rangle}$ are
the mass of electron,
the integrated kinematical factor \cite{doi85,pant}
and the $0\nu\beta\beta$-decay nuclear matrix element,
respectively.  

The present most stringent lower bound on $T_{1/2}^{0\nu-{\rm exp}}$ has been
measured for $^{76}$Ge by Heidelberg-Moscow group and is equal to $5.7\times
10^{25}$ years \cite{bau99,avi}. A condition for obtaining reliable limit for 
fundamental particle physics quantity $\langle m_{\rm ee}\rangle$ is that the
nuclear matrix elements governing the Majorana mass mechanism of
$0\nu\beta\beta$-decay can be calculated correctly \cite{fae98}. However, the
practical nuclear structure calculation always involves some approximations,
which make it difficult to obtain an unambiguous limit on $\langle m_{\rm
ee}\rangle$. We note that the difference between the previous  ($M^{}_{\langle
m_{\rm ee}\rangle} = 4.18$, i.e., ${\langle m_{\rm ee}\rangle} \le 0.18$)
\cite{hax} and the recent more advanced  ($M^{}_{\langle m_{\rm ee}\rangle} =
1.67$, i.e., ${\langle m_{\rm ee}\rangle} \le 0.46$) \cite{cau} shell model
calculations is significant. There is also an other group of nuclear structure
calculations which include the proton-neutron Quasiparticle Random Phase
Approximation  (pn-QRPA) \cite{vog86} and its extensions \cite{pant,si99}.  
Some of them suggest the upper bound on  ${\langle m_{\rm ee}\rangle}$ to be 
in the 0.1 eV range \cite{bau99}.

The aim of the present letter is to discuss the nuclear physics
aspects of the pn-QRPA \cite{vog86}, the renormalized pn-QRPA (pn-RQRPA) 
\cite{toi,simn96} and the
self-consistent pn-RQRPA (pn-SRQRPA) \cite{ds90,bkz00}
calculation of the nuclear matrix element 
$M^{}_{\langle m_{\rm ee}\rangle}$ for $0\nu\beta\beta$-decay of $^{76}$Ge. 
We note that the pn-SRQRPA  results on 
$0\nu\beta\beta$-decay matrix elements have been not presented
in the literature till now.

We  shortly present the main differences between the
above mentioned three QRPA approaches. 

The pn-QRPA has been the most popular theoretical tool 
in description of the 
$\beta$ and $\beta\beta$ decays of medium and heavy open
shell nuclei. However, the pn-QRPA develops a collapse
beyond some critical value of the particle-particle 
interaction strength of nuclear Hamiltonian close to its
realistic value. This phenomena makes predictive power of the obtained
results questionable.  

By implementing the Pauli exclusion principle (PEP) in
an approximate way in the pn-QRPA one gets the 
pn-RQRPA \cite{toi,simn96},
which avoids collapse within a physical range of
particle-particle force and offers more stable solution.
This fact has been  confirmed also within the schematic models.
In addition it was found that by
restoring the PEP a better agreement with the exact 
solution is obtained \cite{schm}. 

The selfconsistent pn-RQRPA (pn-SRQRPA) \cite{jrn80,bkz00,ds90}, 
which is more complex version of the
RQRPA, is becoming increasingly popular to describe strongly
correlated Fermion systems. 
The pn-SRQRPA goes a step further beyond the pn-RQRPA.
In the pn-SRQRPA at the same time the mean field
is changed by minimizing the energy and fixing the number of particles 
in the correlated ground state instead of the uncorrelated BCS one
as it is done in the other versions of the QRPA (pn-QRPA, pn-RQRPA)
\cite{jrn80,bkz00,ds90}.
Thus the pn-SRQRPA is closer to a fully variational theory.

We proceed by writing the expression for the $0\nu\beta\beta$-decay
nuclear matrix element. We note that the contribution to 
$M_{\langle m_{\rm ee}\rangle}$ from the induced pseudoscalar term of the
nucleon current were not considered before. Recently, it
has been found that it is significant  and leads to a modification
of Gamow-Teller and new tensor contributions of
$M_{\langle m_{\rm ee}\rangle}$ \cite{si99}.  Thus $M_{\langle m_{\rm ee}\rangle}$ 
is given as sum of Fermi, Gamow-Teller and tensor contributions
\begin{equation}
M_{\langle m_{\rm ee}\rangle} = -\frac{M_F}{g^2_A} + M_{GT} + M_{T}
\label{eq:3}   
\end{equation}
with $g_A = 1.25$. Expressed in relative 
coordinates and using the second quantization formalism  
$M_{\langle m_{\rm ee}\rangle}$ takes the form
\begin{eqnarray}
\label{eq:4}
M_{\langle m_{\rm ee}\rangle} & = &
\sum_{J^{\pi}} \sum_{{{p n p' n' } \atop m_i m_f {\cal J}  }}
(-)^{j_{n}+j_{p'}+J+{\cal J}}(2{\cal J}+1)
\left\{\matrix{
j_p & j_n & J \cr
j_{n'} & j_{p'} & {\cal J}}\right\} \nonumber \\
&& \times
\langle p(1), p'(2);{\cal J}|f(r_{12})\tau_1^+ \tau_2^+ 
{\cal O}_{\langle m_{\rm ee}\rangle} (12)
f(r_{12})|n(1) ,n'(2);{\cal J}\rangle \nonumber \\
&& \times
\langle 0_f^+ \parallel
\widetilde{[c^+_{p'}{\tilde{c}}_{n'}]_J} \parallel J^\pi m_f\rangle
\langle J^\pi m_f|J^\pi m_i\rangle
\langle J^\pi m_i \parallel [c^+_{p}{\tilde{c}}_{n}]_J \parallel
0^+_i\rangle.
\end{eqnarray}
Here, $f(r_{12})$ is the short-range correlation function 
and ${\cal O}_{\langle m_{\rm ee}\rangle}(12)$ 
represents the coordinate and spin dependent part of the
two-body $0\nu\beta\beta$-decay transition operator
\begin{equation}
{\cal O}_{\langle m_{\rm ee}\rangle} (12) = 
- \frac{H_{F}(r_{12})}{g^2_A}   + 
H_{GT}(r_{12}) {\bf \sigma}_{12} +
H_{T}(r_{12}) {\bf S}_{12},
\label{eq:5}
\end{equation}
where  $\sigma_{12} = {\vec{ \sigma}}_1\cdot {\vec{ \sigma}}_2$,
$S_{12} = 3({\vec{ \sigma}}_1\cdot \hat{q}{\vec{\sigma}}_2 \cdot \hat{q})
- \sigma_{12},$ and
\begin{equation}
\label{eq:6}   
H_{K}(r_{12}) =\frac{2}{\pi g^2_A}
\frac{R}{r_{12}} \int_{0}^{\infty} 
\frac{\sin(qr_{12})} {q+E^m(J)- (E^i + E^f)/2}
h_{K}(q^2)\,d{q}, (K = F, GT, T) 
\end{equation}
with
\begin{eqnarray}
h_{F}(q^2) &=& g^2_V (q^2) g^2_A, \nonumber \\
h_{GT}(q^2) & = & 
g^2_A (q^2) +\frac{1}{3}\frac{g^2_P (q^2) q^4}{4 m^2_p}
-\frac{2}{3}\frac{g_A (q^2) g_P (q^2) q^2}{2 m_p}, \nonumber \\
h_{T}(q^2) & = & 
\frac{2}{3}\frac{g_A (q^2) g_P (q^2) q^2}{2 m_p}
- \frac{1}{3}\frac{g^2_P (q^2) q^4}{4 m^2_p}.
\label{eq:7}   
\end{eqnarray}
Here, $R$ is the mean nuclear radius \cite{si99}. 
$E^i$, $E^f$ and $E^{m}(J)$ are respectively the energies of the 
initial, final and intermediate nuclear state with angular momentum  $J$.
The momentum dependence of the vector, axial-vector and pseudoscalar
formfactors ($g_V(q^2)$, $g_A(q^2)$ and $g_P(q^2)$) is given 
in Ref. \cite{si99}.

We note that the first sum on the r.h.s of Eq. (\ref{eq:4}) represents
summation over all multipolarities. 
The form of the one-body transition densities [see Eq. (\ref{eq:4})]
to excited states
$|J^\pi m_i\rangle$ and $|J^\pi m_f\rangle$ generated from  the initial (A,Z) 
and the final (A,Z+2) ground states $|0^+_i\rangle$ and $0^+_f\rangle$
within the pn-RQRPA and the pn-SRQRPA is the same and can be 
found together with other details of the nuclear structure
in Refs. \cite{simn96,si99,bkz00}. The difference consists only 
 in the calculated value 
of the renormalized coefficients $D$, which in the case of the
pn-QRPA is just equal to unity. The overlap factor entering the
expression (\ref{eq:4}) can be find in Ref. \cite{simr88}.

The calculation of   $0\nu\beta\beta$-decay matrix elements
of $^{76}$Ge are performed within two model spaces
both for protons and neutrons as follows:
i) The model space I (m.s. I) 
consists of the full $3-4\hbar\omega$ major oscillator shells
and has been considered in the pn-QRPA studies of Ref. \cite{vog86}
(9-levels model space).
ii) The model space II (m.s. II)
comprises the full $2-5\hbar\omega$ major shells (12 levels 
model space).

The single particle energies
were  obtained by using a  Coulomb--corrected Woods--Saxon
potential with Bertsch parameterization. Two-body G-matrix elements
were calculated from the Bonn one-boson exchange potential within 
within the Brueckner theory. In the pn-QRPA and 
the pn-RQRPA approaches
pairing interactions have been adjusted to fit 
the empirical pairing gaps. In the pn-SRQRPA approach the
pairing matrix elements of the NN interaction have not been
rescaled as the mean field is directly related to the excited
states.  The  particle-particle and particle-hole channels of 
the G-matrix interaction of the nuclear Hamiltonian $H$ 
are renormalized by introducing
the parameters $g_{\rm pp}$ and $g_{\rm ph}$, respectively. 

\begin{figure}
\vspace*{-4.0cm}
\centerline{\epsfig{file=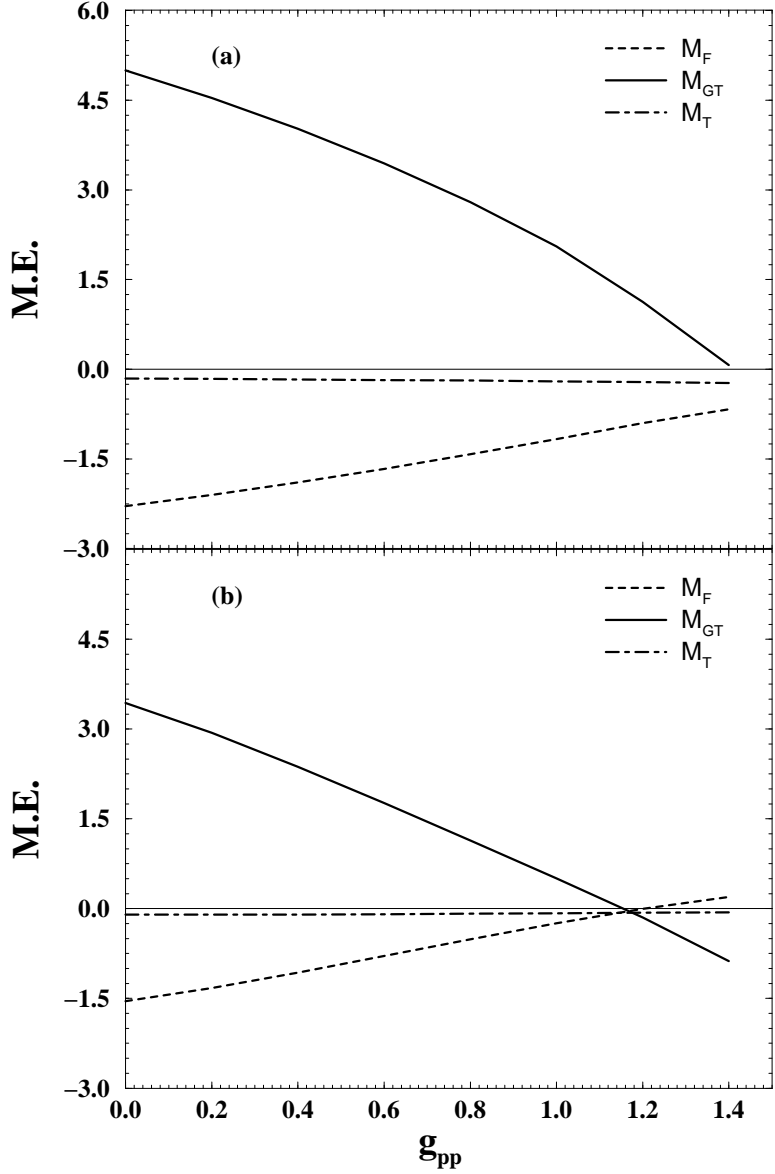,height=24.cm}}
\vspace{-3.0cm}
\caption{
The Fermi ($M_{F}$, dashed line), the Gamow-Teller 
($M_{GT}$, solid line) and tensor 
($M_{T}$, dot-dased line) nuclear matrix elements 
for the $0\nu\beta\beta$-decay of $^{76}$Ge 
calculated within the pn-RQRPA (a) and  pn-SRQRPA (b)  are plotted as  a
function of the particle-particle coupling constant $g_{\rm pp}$.
The 12-level model space II (the full $2-4\hbar\omega$ major oscillator shells)
was assumed and $g_{\rm ph}=1.0$
}
\label{fig.1}
\end{figure}

In Fig. \ref{fig.1} 
the calculated partial matrix elements $M_F$, $M_{GT}$
and $M_T$ are plotted as function of $g_{\rm pp}$  ($g_{\rm ph}=1.0$) for
the pn-RQRPA and pn-SRQRPA approaches. The larger 12-levels model space
is considered. One find a strong dependence of $M_F$ and $M_{GT}$ 
on $g_{\rm pp}$. The smallest $M_T$ matrix element is rather insensitive
to this parameter. In general, the behavior of plotted matrix elements
is similar for both approaches. Nevertheless, the Fermi and GT 
matrix elements  of the pn-SRQRPA one reach zero value inside  
the physical range of $g_{\rm pp}$ parameter ($0.8 \le g_{\rm pp} \le 1.2$).

\begin{figure}
\vspace*{-3.0cm}
\centerline{\epsfig{file=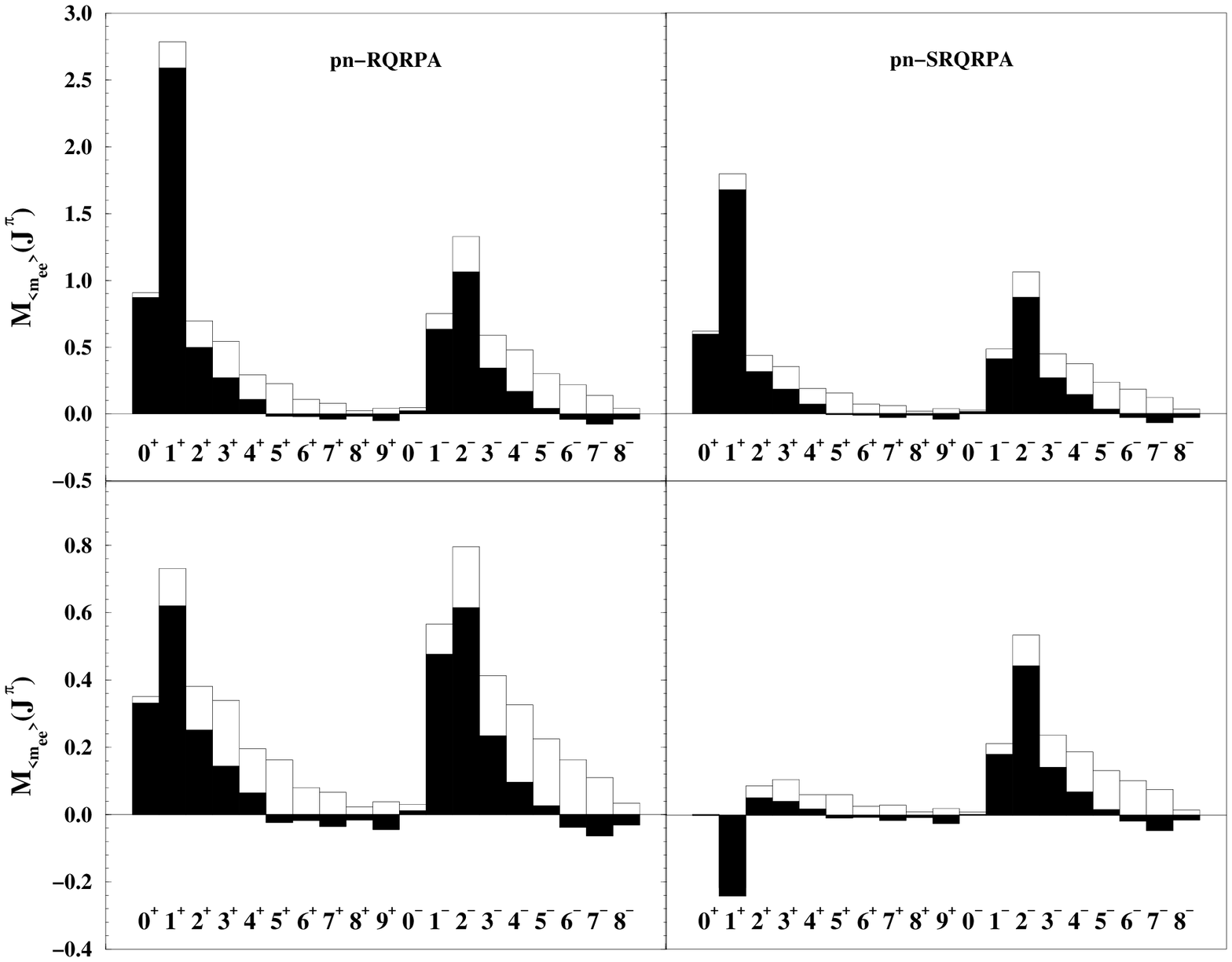,height=24.cm}}
\vspace{-4.5cm}
\caption{
The contributions of different multipolarities to the nuclear matrix element
$M_{\langle m_{\rm ee}\rangle}$ for the $0\nu\beta\beta$-decay of $^{76}$Ge 
calculated within the pn-RQRPA (left figures) and pn-SRQRPA 
(right figures) without (open bar) and
with (closed bar) consideration of the nucleon short-hand correlations 
(s.r.c.). The model space is the same as in Fig. 1 and $g_{\rm ph}=1.0$. 
The results in upper and lower graphs were obtained 
for $g_{\rm pp}=0.0$ and $g_{\rm pp}=1.0$, respectively. 
}
\label{fig.2}
\end{figure}

In Fig. \ref{fig.2} 
the multipole decomposition (according to intermediate multipoles
$J^\pi$) of matrix element $M_{\langle m_{\rm ee}\rangle}$  for
the pn-RQRPA and pn-SRQRPA approaches and two values of $g_{\rm pp}$
($g_{\rm pp} = 0.0$ and $g_{\rm pp} = 1.0$) is presented. The model space 
and $g_{\rm ph}$ is the same as in Fig. \ref{fig.1}. 
The filled and open bars represent the calculations with and without
consideration of the two--nucleon short--range correlations (s.r.c.),
respectively. One clearly sees that the s.r.c. suppress the contributions
from higher multipolarities. It just proves that the Majorana neutrino
mass exchange mechanism is predominantly long-range interaction in the
nucleus. We notice also that the contributions associated with 
even parity multipolarities are more
influenced by the change of $g_{\rm pp}$ (especially in the case of
the pn-SRQRPA) as those associated with the 
 negative ones. The differences between the even parity multipole 
contributions calculated via the pn-RQRPA and the 
pn-SQRPA approaches is remendeous for $g_{\rm pp}=1.0$. 
This effect can be attributed to the different structure of the 
mean field in both approaches.

\begin{figure}
\vspace*{-3.0cm}
\centerline{\epsfig{file=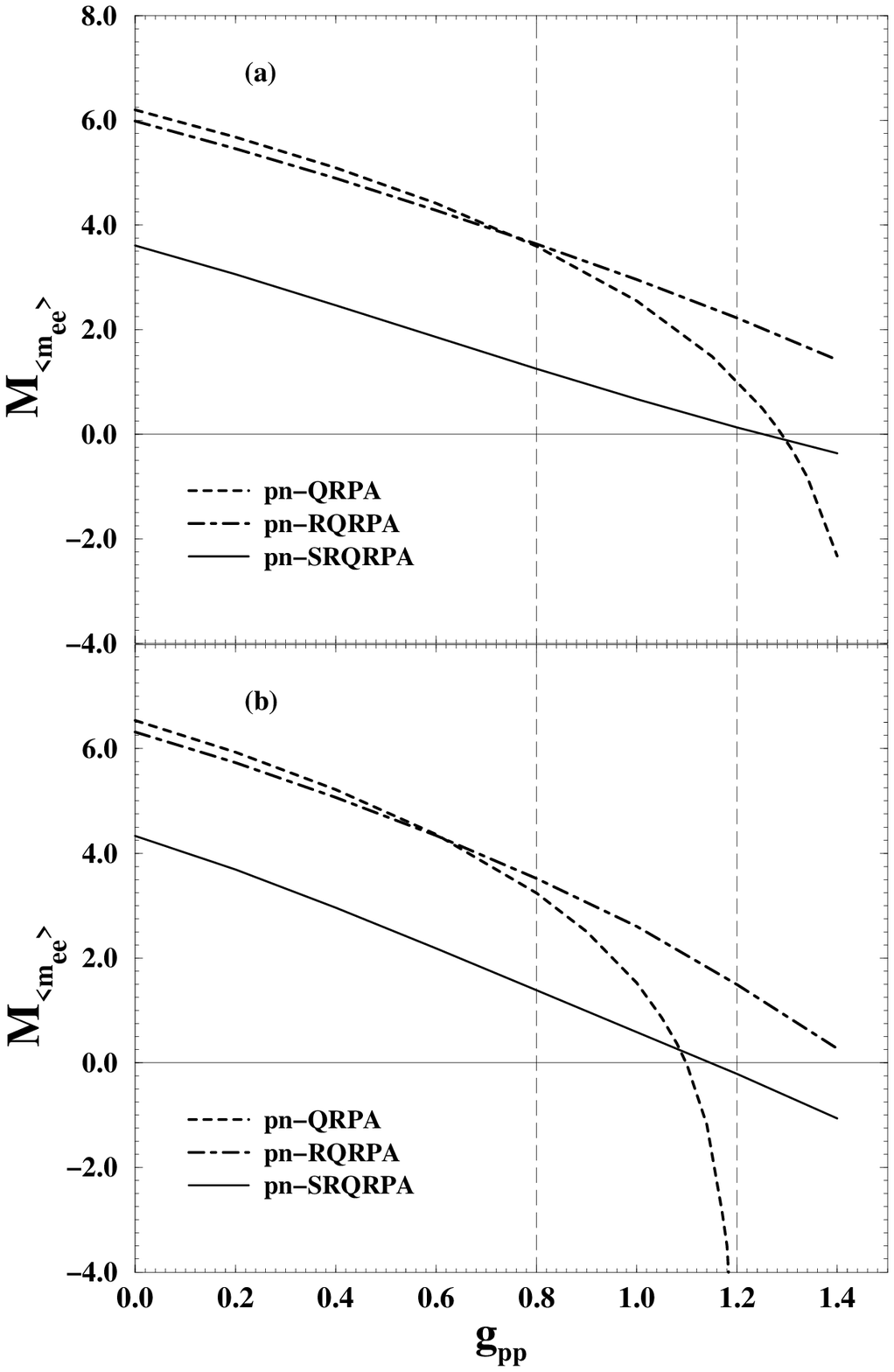,height=24.cm}}
\vspace{-2.0cm}
\caption{ The nuclear matrix element
$M_{\langle m_{\rm ee}\rangle}$ for the $0\nu\beta\beta$-decay of $^{76}$Ge 
calculated within the pn-QRPA (dashed line), 
pn-RQRPA (dot-dashed line)  and pn-SQRPA (solid line)
in the framework of the 9-level model space (a)
(the full $3-4\hbar\omega$ major oscillator shells),
and the 12-level model space (b)
(the full $2-4\hbar\omega$ major oscillator shells)
is plotted as  a
function of the particle-particle coupling constant $g_{\rm pp}$.
$g_{\rm ph}$ was taken to be unity.
}
\label{fig.3}
\end{figure}

In Fig. \ref{fig.3} 
the nuclear matrix elements $M_{\langle m_{\rm ee}\rangle}$  as function of 
particle--particle strength $g_{\rm pp}$ for all three QRPA
approaches (pn-QRPA, pn-RQRPA, pn-SRQRPA) is shown. One notices
that the pn-RQRPA and the pn-SRQRPA results are much less sensitive
to the truncation of model space and $g_{\rm pp}$ as the pn-QRPA
ones. The unpleasant feature of the pn-QRPA and the pn-SRQRPA is that
for the larger model space $M_{\langle m_{\rm ee}\rangle}$  crosses zero within the 
physical interval of $g_{\rm pp}$ what increase uncertainty in respect
to the extracted value on $\langle m_{\rm ee}\rangle$ from non-observability of
$0\nu\beta\beta$-decay considerably.

The real value of the $g_{\rm pp}$ parameter is expected to be close
to unity. In Table \ref{tabl.1} we present 
$M_{\langle m_{\rm ee}\rangle}$ for this value of $g_{\rm pp}$ parameter together with 
the corresponding upper bound
on $\langle m_{\rm ee}\rangle$ deduced from the Heisenberg-Moscow experiment.
One notices that only the pn-QRPA results are sensitive
to the truncation of model space. The new calculated pn-SRQRPA results predict
considerably larger value for upper limit on  $\langle m_{\rm ee}\rangle$ in comparison
with the pn-QRPA and the pn-RQRPA ones. It is due to the mean-field 
modifications by the number self-consistent approach which use
to diminish the pairing gap \cite{mahi00}.  

In summary, the nuclear matrix element associated with the light Majorana
neutrino mass mechanism of $0\nu\beta\beta$-decay of $^{76}$Ge has been
calculated in the selfconsistent pn-RQRPA approach which received much
attention recently. This approach predicts considerably suppressed
$0\nu\beta\beta$-decay matrix element, what, when confronted with the present
experimental data, leads to a less stringent limit on the effective Majorana
neutrino mass parameter $\langle m_{\rm ee}\rangle$. The reduction of the
matrix element is due to the modification of the mean field by the combined
selfconsistent BCS+RQRPA approach which influences mostly the even parity
multipole contributions.

This work was supported in part by the State Committee for Scientific 
Researches (Poland), Contract No. 2 P03 B00516.

%%%%%%%%%%%%%%%%%%%%%%%%%%%%%%%%%%%%%%%%%%%%%%%%%%%%%%%%%%%%%%%%%%%%%%%%
%%%%%%%%%%%%%%%%%           Table section           %%%%%%%%%%%%%%%%%%%% 
%%%%%%%%%%%%%%%%%%%%%%%%%%%%%%%%%%%%%%%%%%%%%%%%%%%%%%%%%%%%%%%%%%%%%%%%

\begin{table}[h]
\caption{Calculated nuclear matrix element $M_{\langle m_{\rm ee}\rangle}$ 
for $0\nu\beta\beta$-decay of $^{76}$Ge 
within the pn-QRPA, pn-RQRPA and the pn-SRQRPA
and the corresponding upper limit on the effective neutrino mass
$\langle m_{\rm ee}\rangle$ deduced from the experimental lower bound for a 
half--life $T^{0\nu}_{1/2} (^{76}Ge) \ge 5.7\times 10^{25}$ years
\protect\cite{bau99}.
Here, m.s. I and m.s. II denote the model spaces which comprice
the full $3-4\hbar\omega$ and $2-4\hbar\omega$ 
major oscillator shells, respectively.
}
\begin{tabular}{cccccccc}
 &  & \multicolumn{2}{c}{pn-QRPA} & \multicolumn{2}{c}{pn-RQRPA} 
& \multicolumn{2}{c}{pn-SRQRPA} \\
\cline{3-4} \cline{5-6} \cline{7-8}
  & $g_{\rm ph}$ & m.s. I & m.s. II & m.s. I & m.s. II & m.s. I & m.s. II 
\\ \hline
$M_{\langle m_{\rm ee}\rangle}$ & 0.8 & 2.76 & 1.66 & 3.15 & 2.80 & 0.72 & 
0.65 \\
$\langle m_{\rm ee}\rangle$             &     & 0.28 & 0.46 & 0.24 & 0.27 & 
1.1  & 1.2 \\
 & & & & &  \\
$M_{\langle m_{\rm ee}\rangle}$ & 1.0 & 2.54 & 1.53 & 2.95 & 2.61 & 0.66 & 
0.59 \\
$\langle m_{\rm ee}\rangle$             &     & 0.30 & 0.50 & 0.26 & 0.29 & 
1.2  & 1.3 \\
\end{tabular}
\label{tabl.1}
\end{table}

%%%%%%%%%%%%%%%%%%%%%%%%%%%%%%%%%%%%%%%%%%%%%%%%%%%%%%%%%%%%%%%%%%%%%%%%
%%%%%%%%%%%%%%%%%          Figures section          %%%%%%%%%%%%%%%%%%%% 
%%%%%%%%%%%%%%%%%%%%%%%%%%%%%%%%%%%%%%%%%%%%%%%%%%%%%%%%%%%%%%%%%%%%%%%%

\end{document}